\documentstyle[aps,psfig,multicol]{revtex}
\input{epsf}
\newcommand{\bcols}{\ifpreprintsty\else\begin{multicols}{2}\fi}
\newcommand{\ecols}{\ifpreprintsty\else\end{multicols}\fi}
\begin{document}
\draft
\title{Golden rule decay versus Lyapunov decay of the quantum Loschmidt echo}
\author{Ph. Jacquod$^1$, P.G. Silvestrov$^{1,2}$, and C.W.J. Beenakker$^1$}
\address{$^1$ Instituut-Lorentz, Universiteit Leiden, P.O. Box 9506, 2300 RA
Leiden, The Netherlands\\
$^2$ Budker Institute of Nuclear Physics, 630090 Novosibirsk, Russia}
\date{August 2001}
\maketitle
\begin{abstract}
The overlap of two wave packets evolving in time with slightly different
Hamiltonians decays exponentially $\propto e^{-\gamma t}$, for perturbation
strengths $U$ greater than the level spacing $\Delta$. We present numerical
evidence for a dynamical system that the decay rate $\gamma$ is given by the
{\em smallest\/} of the Lyapunov exponent $\lambda$ of the classical chaotic
dynamics and the level broadening $U^{2}/\Delta$ that follows from the golden
rule of quantum mechanics. This implies the range of validity
$U>\sqrt{\lambda\Delta}$ for the perturbation-strength independent decay rate
discovered by Jalabert and Pastawski [Phys.\ Rev.\ Lett.\ {\bf 86}, 2490
(2001)].
\end{abstract}
\bcols
The search for classical Lyapunov exponents in quantum mechanics is a
celebrated problem in quantum chaos \cite{Per93}. Motivated by NMR experiments
on spin echoes \cite{Pas00}, Jalabert and Pastawski \cite{Jal01} have given
analytical evidence, supported by computer simulations \cite{Cuc01}, that the
Lyapunov exponent governs the time dependence of the fidelity
\begin{equation}
M(t)=|\langle\psi|\exp(iHt)\exp(-iH_{0}t)|\psi\rangle|^{2} \label{Mdef}
\end{equation}
with which a wave packet $\psi$ can be reconstructed by inverting the
dynamics with a perturbed Hamiltonian $H=H_{0}+H_{1}$. They have called this
the problem of the ``quantum Loschmidt echo''. The fidelity $M(t)$ can
equivalently be interpreted as the decaying overlap of two wave functions that
start out identically and evolve under the action of two slightly different
Hamiltonians, a problem first studied in perturbation theory by Peres
\cite{Per84}.

Perturbation theory breaks down once a typical matrix element $U$ of $H_{1}$
connecting different eigenstates of $H_0$
becomes greater than the level spacing $\Delta$. Then the eigenstates of $H$,
decomposed into the eigenstates of $H_{0}$, contain a large number of
non-negligible components. The distribution $\rho(E)$ (local spectral density)
of these components over energy has a Lorentzian form 
\begin{equation}
\rho(E)=\frac{\Gamma}{2 \pi (E^2+\Gamma^2/4)},
\end{equation}
with a spreading width
$\Gamma\simeq U^{2}/\Delta$ given by the golden rule \cite{Jac95,Kota01}. 
A simple
calculation in a random-matrix model gives an average decay
$\bar{M}\propto\exp(-\Gamma t)$ governed by the same golden rule width. 
This should
be contrasted with the exponential decay $\bar{M}\propto\exp(-\lambda t)$
obtained by Jalabert and Pastawski \cite{Jal01}, which is governed by the
Lyapunov exponent $\lambda$ of the classical chaotic dynamics.

Since the random-matrix model has by construction an infinite Lyapunov
exponent,  one way to unify both results would be to have an exponential decay
with a rate set by the smallest of $\Gamma$ and $\lambda$. We will in what
follows present numerical evidence for this scenario, using a dynamical system
in which we can vary the relative magnitude of $\Gamma$ and $\lambda$. There
exists a third energy scale, the inverse of the Ehrenfest time $\tau_{\rm E}$,
that is smaller than the Lyapunov exponent by a factor logarithmic in the
system's effective Planck constant. In our numerics we do not have 
enough orders of
magnitude between $1/\tau_{\rm E}$ and $\lambda$ to distinguish between the
two, so that our findings remain somewhat inconclusive in this respect.

Because $\Gamma$ cannot become bigger than the band width $B$ of $H_{0}$
(we are interested in the regime $H_1<H_0$), a
consequence of a decay $\bar{M}\propto\exp[-t\,{\rm min}\,(\lambda,\Gamma)]$ 
is
that the regime of Lyapunov decay can only be reached with increasing $U$ if
$\lambda$ is considerably less than $B$. That would exclude typical fully
chaotic systems, in which $\lambda$ and $B$ are comparable, and set limits
of observability of the Lyapunov decay.

The crossover from the golden rule regime to a regime with a
perturbation-strength independent decay, obtained here for the Loschmidt
echo, should be distinguished from the corresponding crossover in the
local spectral density $\rho(E)$, obtained by Cohen and Heller
\cite{Cohen00}. The Fourier transform of $M(t)$ would be equal to $\rho(E)$ if
$\psi$ would be an eigenstate of $H_{0}$ rather than a wave packet. The
choice of a wave packet instead of an eigenstate does not matter in
the golden rule regime, but is essential for a decay rate
given by the Lyapunov exponent.

The dynamical model that we have studied is the kicked top \cite{Haa00}, with
Hamiltonian
\begin{equation}
H_{0}=(\pi/2\tau)S_{y}+(K/2S)S_{z}^{2}\sum_{n}\delta(t-n\tau). \label{H0def}
\end{equation}
It describes a vector spin (magnitude $S$) that undergoes a free precession
around the $y$-axis perturbed periodically (period $\tau$) by sudden rotations
around the $z$-axis over an angle proportional to $S_{z}$. The time evolution
of a state after $n$ periods is given by the $n$-th power of the Floquet
operator
\begin{equation}
F_{0}=\exp[-i(K/2S)S_{z}^{2}]\exp[-i(\pi/2)S_{y}]. \label{Fdef}
\end{equation}
Depending on the kicking strength $K$, the classical dynamics is regular,
partially chaotic, or fully chaotic. The dependence of the Lyapunov exponent
$\lambda$ on $K$ is plotted in the inset to Fig.\ 1 (cf.\ Ref.\ \cite{Ben76}).
The error bars reflect the spread in $\lambda$ in different regions of phase
space, in particular the presence of islands of stability. 
For $K\gtrsim 9$ the error bars vanish because the system has become
fully chaotic. For the reversed time evolution we introduce as a 
perturbation a
periodic rotation of constant angle around the $x$-axis, slightly
delayed with respect to the kicks $H_0$,
\begin{equation}
H_{1}=\phi S_{x}\sum_{n}\delta(t-n\tau-\epsilon). \label{H1def}
\end{equation}
The corresponding Floquet operator is $F=\exp(-i\phi S_{x})F_{0}$. We have set
$\hbar=1$ and in what follows we will also set $\tau=1$ for
ease of notation.

Both $H$ and $H_{0}$ conserve the spin magnitude. We choose the initial wave
packets as coherent states of the spin SU(2) group \cite{Per86}, i.e.\ states
which minimize the Heisenberg uncertainty in phase space (in our case on a
sphere of fixed radius) at the effective Planck constant
$h_{\rm eff} \sim S^{-1}$. The corresponding Ehrenfest time
is $\tau_{\rm E}=\lambda^{-1}\ln S$ \cite{Haa92}. We take $S=500$ and average
$M(t=n)=|\langle\psi|(F^{\dagger})^{n}F_{0}^{n}|\psi\rangle|^{2}$ over 100
initial coherent states $\psi$.

\begin{figure}
\vspace{8mm}
\epsfxsize=3.1in
\epsfysize=2.3in
\epsffile{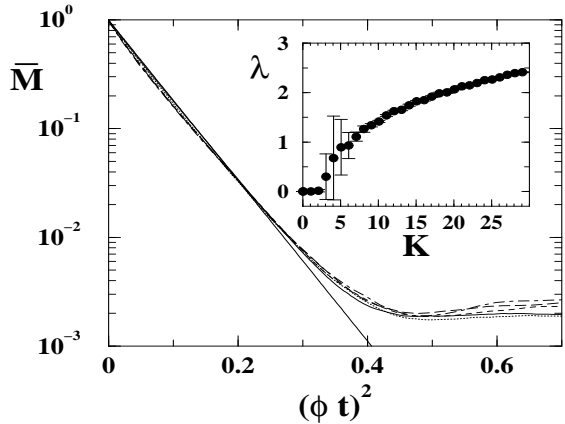}

\vspace{2 mm}

\caption
{Decay of the average fidelity $\bar{M}$ for the quantum kicked top 
with $K=13.1$ and $S=500$, as a function of the squared
rescaled time $(\phi t)^2$. The perturbation
strengths range between $\phi=10^{-7}$ and $10^{-6}$. 
The straight line corresponds to the Gaussian decay (\ref{perturbation})
valid in the perturbative regime. Inset : Numerically
computed Lyapunov exponent for the classical kicked top as a function of the 
kicking strength $K$. Dots correspond to averages taken over
$10^4$ initial conditions (see Ref. \protect\cite{Ben76}).
The error bars reflect different results 
obtained with different initial conditions. The vanishing of error bars
indicates the disappearance of islands of regular dynamics.}
\label{fig:fig1}
\end{figure}

We first show results in the fully chaotic regime $K>9$, 
where we choose the initial
states randomly over the entire phase space. The local spectral density 
$\rho(\alpha)$ of the
eigenstates of $F$ (in the basis of the eigenstates of $F_{0}$ with 
eigenphases $\alpha$) is plotted for three different $\phi$'s 
in the inset to Fig. 2. The curves can be fitted
by Lorentzians from which we extract the spreading width $\Gamma$. (It is
given up to numerical coefficients by $\Gamma\simeq U^{2}/\Delta$, $U\simeq\phi
\sqrt{S}$, $\Delta\simeq 1/S$). The golden rule regime $\Gamma\gtrsim\Delta$ is
entered at $\phi_{c}\approx 1.7\cdot 10^{-4}$. For $\phi\ll\phi_{c}$ we are in
the perturbative regime, where eigenstates of $F$ do not appreciably
differ from
those of $F_0$ and eigenphase differences can be calculated in first
order perturbation theory.
We then expect the Gaussian decay
\begin{equation}
\bar{M}\propto\exp(-U^{2}t^{2})\Rightarrow\ln\bar{M}\propto (\phi t)^{2}.
\label{perturbation}
\end{equation}
This decay is evident in Fig.\ 1, which shows 
$\bar{M}$ as a function
of $(\phi t)^{2}$ on a semilogarithmic scale for $\phi \le 10^{-6}$.
The decay (\ref{perturbation}) stops when $\bar{M}$ approaches
$M_{\infty} = 1/2S$, being the inverse of the dimension of the
Hilbert space. 
This saturation reflects the finiteness of the system and eventually
prevails at long times independently of the strength of the perturbation.

\begin{figure}
\vspace{8mm}
\epsfxsize=3.1in
\epsfysize=2.3in
\epsffile{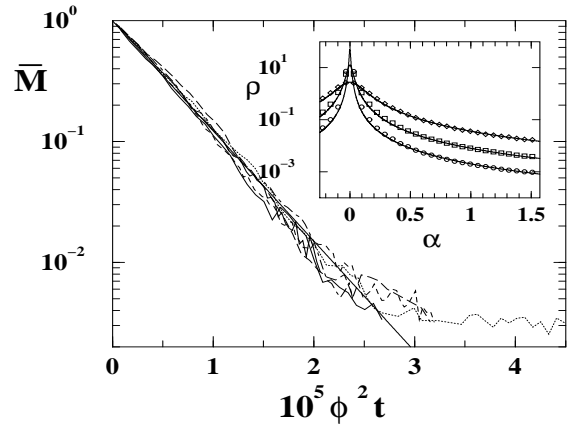}

\vspace{2 mm}

\caption
{Decay of $\bar{M}$ 
in the golden rule regime for
kicking strengths $K=13.1$, 17.1, 21.1
as a function of the rescaled time $\phi^{2} t$. Perturbation 
strengths range from $\phi=10^{-4}$ to $10^{-3}$.
Inset: Local spectral density of states 
for $K=13.1$ and perturbation
strengths $\phi=2.5 \cdot 10^{-4}, 5 \cdot 10^{-4}$, $10^{-3}$. 
The solid curves are Lorentzian fits, from which the
decay rate $\Gamma \approx 0.84\, \phi^2 S^2$ is extracted.
The solid line in the main plot gives the decay $\bar{M} \propto
\exp(-\Gamma t)$ with this value of $\Gamma$.}
\label{fig:fig2}
\end{figure}

For $\phi > \phi_c$ one enters
the golden rule regime, where the Lorentzian 
spreading of eigenstates of $F$ over those of $F_0$ results in
the exponential decay 
\begin{equation}
\bar{M}\propto\exp(-U^{2}t/\Delta)\Rightarrow\ln\bar{M}\propto \phi^{2} t.
\label{goldenrule}
\end{equation}
The data presented in Fig.\ 2 clearly confirm the validity 
of the scaling (\ref{goldenrule}). There is no dependence 
of $\bar{M}$ 
on $K$ in this regime of moderate (but non-perturbative) values of $\phi$,
i.e. no dependence on the Lyapunov exponent
($\lambda$ varies by a factor of 1.4 for the different values of $K$
in Fig.\ 2).

We cannot satisfy $\lambda<\Gamma$ in the fully chaotic regime, for the reason
mentioned in the introduction: The band width $B$ (which is an upper limit for
$\Gamma$) is $B=\pi/2$ (in units of $1/\tau$), while $\lambda\gtrsim 1$ for
fully developed chaos in the kicked top (see the inset to Fig.\ 1). 
For this reason, when the perturbation strength $\phi$ is further
increased, the golden rule decay rate saturates at the band width ---
before reaching the Lyapunov exponent. This is shown
in Fig.\ 3. There is no trace of a Lyapunov decay in this fully chaotic
regime.

\begin{figure}
\vspace{8mm}
\epsfxsize=3.1in
\epsfysize=2.7in
\epsffile{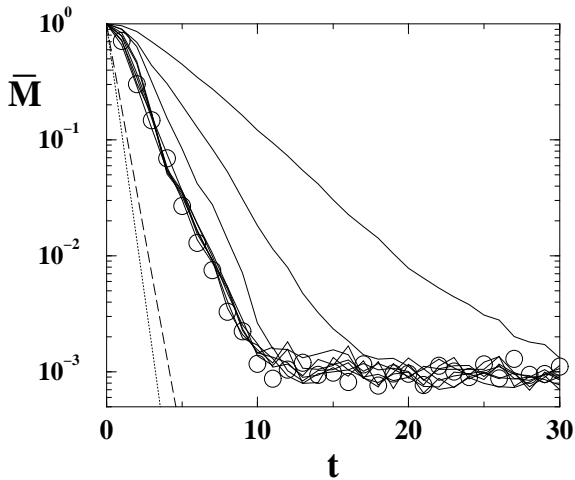}

\vspace{2 mm}

\caption
{Decay of $\bar{M}$ in the golden rule regime without rescaling
of time, for $K=13.1$,
$\phi = j \cdot 10^{-3}$, ($j$=1, 1.5, 2, \ldots5) (solid curves)
and $K=21.1$, $\phi=3 \cdot 10^{-3}$ (circles). Dashed
and dotted lines show exponential decays with Lyapunov
exponents $\lambda = 1.65$ and 2.12, corresponding to 
$K=13.1$ and 21.1, respectively. The decay slope saturates at 
$\phi \approx 2.5 \cdot 10^{-3}$, when $\Gamma$ reaches the bandwidth.}
\label{fig:fig3}
\end{figure}

We therefore reduce $K$ to values in the range $2.7 \le K \le 4.2$, 
which allows us to vary the Lyapunov exponent over a wider range 
between 0.22 and 0.72. In this range the 
classical phase space is mixed and we have coexisting regular and 
chaotic trajectories. We choose 
the initial coherent states in the chaotic region (identified
numerically through the participation ratio). Because
the chaotic region still occupies more than 80\% of the
phase space for the smallest value of $K$ considered,
nonuniversal effects (e.g. nonzero overlap of
our initial wavepackets with regular eigenfunctions of $F_0$ or $F$)
should be negligible. We expect a
crossover from the golden rule decay (\ref{goldenrule}) to the
Lyapunov decay \cite{Jal01}
\begin{equation}
\bar{M}\simeq\exp(-\lambda t)\Rightarrow\ln\bar{M}\propto\lambda t,
\label{Lyapunov}
\end{equation}
once $\Gamma$ exceeds $\lambda$. This expectation is borne out by our
numerical simulations, see Fig.\ 4.

In conclusion, we have presented numerical evidence for the existence of three
distinct regimes of exponential decay of the Loschmidt echo: the perturbative
regime (\ref{perturbation}), the golden rule regime (\ref{goldenrule}), and the
Lyapunov regime (\ref{Lyapunov}). The perturbation strength independent decay
in the Lyapunov regime is reached in our simulation if $\lambda<\Gamma$, which
prevents its occurence for fully developed chaos in the model
considered here. Our numerics are limited by a
relatively small window between $\lambda$ and $1/\tau_{\rm E}$ (a factor $\ln
S\approx 6$). It remains to be seen if the Lyapunov 
decay can be observed under
conditions of fully developed chaos and $\Gamma < \lambda$
by increasing $S$ so that $1/\tau_{\rm E}$
becomes larger than $\Gamma$. It is noteworthy that
for a Lyapunov decay $\bar{M} \propto \exp(-\lambda t)$, 
the saturated fidelity $M_{\infty}=1/2S$ 
is reached at the Ehrenfest time $\tau_{\rm E}$ (as can also be seen in 
Fig.\ 4), 
so that a Lyapunov decay for $t \lesssim \tau_{\rm E}$ rules out
golden rule decay for later times.
Similar investigations in strongly chaotic systems with small Lyapunov
exponents (like the Bunimovich stadium with short straight segments)
are highly desirable. 

\begin{figure}
\vspace{8mm}
\epsfxsize=3.1in
\epsfysize=2.3in
\epsffile{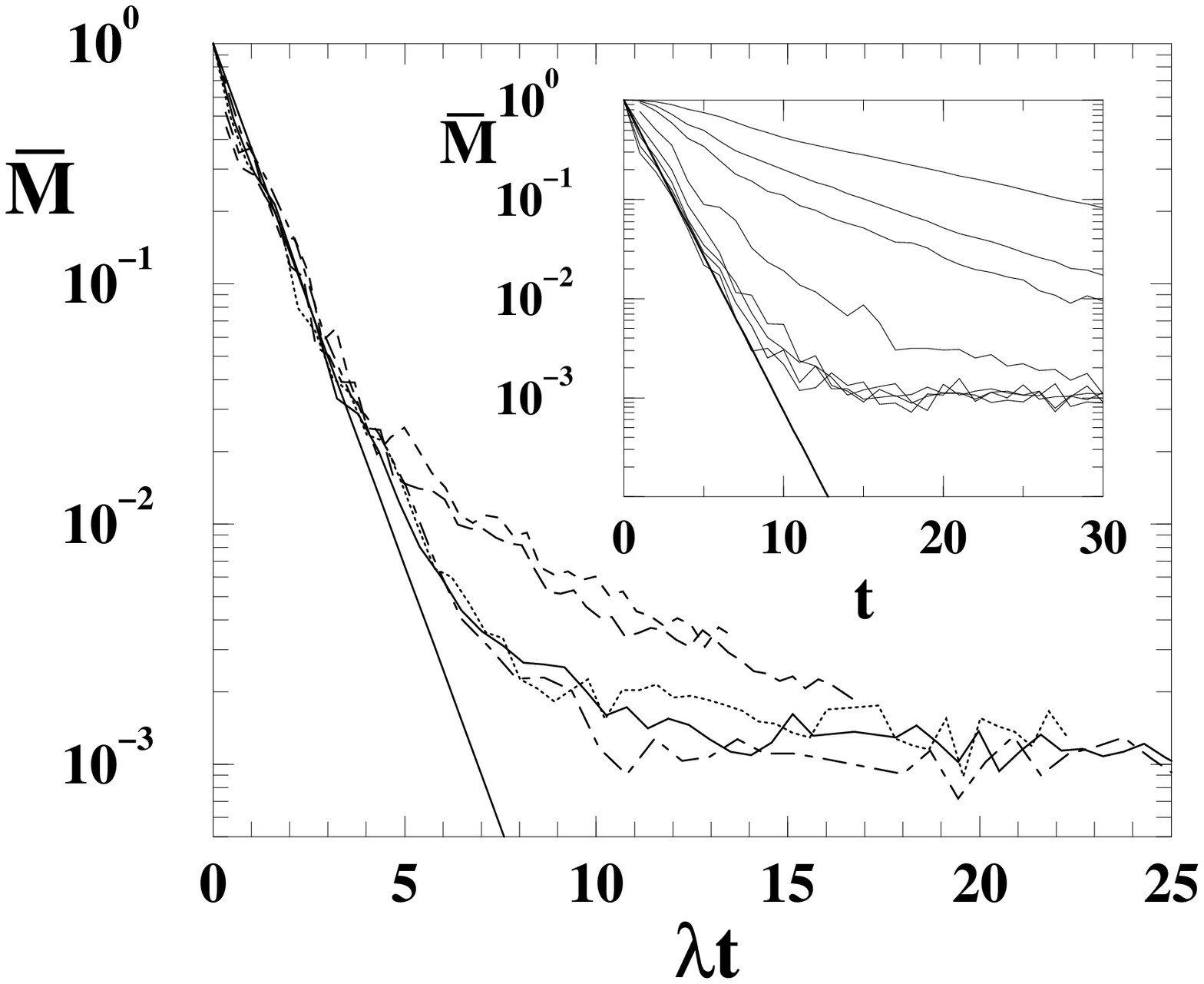}

\vspace{2 mm}

\caption
{Decay of $\bar{M}$ in the Lyapunov regime,
for $\phi=2.1 \cdot 10^{-3}$, $K=2.7$, 3.3, 3.6, 3.9, 4.2.
The time is rescaled with the Lyapunov exponent
$\lambda$, ranging from 0.22 to 0.72. The straight solid line indicates the 
decay $\bar{M} \propto \exp(-\lambda t)$. 
Inset: $\bar{M}$ for $K=4.2$ and different
$\phi=j \cdot 10^{-4}$, $j=$1, 2, 3, 4, 5, 9, 17, 25. 
The decay slope saturates at the value $\phi \approx 1.7 \cdot 10^{-3}$
for which $\Gamma \approx \lambda$, even though $\Gamma$ 
keeps on increasing. This demonstrates the decay law $\bar{M} \propto
\exp(-\gamma t$ with $\gamma = {\rm min}(\Gamma,\lambda)$.}
\label{fig:fig4}
\end{figure}

This work was supported by the Swiss National Science Foundation and by the
Dutch Science Foundation NWO/FOM. We acknowledge helpful comments from 
D. Cohen, F. Haake, and R. A. Jalabert.

\ecols
\end{document}